\documentclass[12pt]{article}
\usepackage{amssymb}

\newcommand{\be}{\begin{equation}}
\newcommand{\ee}{\end{equation}}

\newcommand{\lmmcs }{{\cal L}_{M+MCS}}

\newcommand{\lm}{{\cal L}_{{\rm master}}}
\newcommand{\lmsd}{{\cal L}_{M+SD}}
\newcommand{\jo}{J_{\nu}^{(0)}}
\newcommand{\p}{\partial}

\newcommand{\nn}{\nonumber}
\newcommand{\emn}{\epsilon_{\mu\nu\gamma}\partial^{\gamma}}

\newcommand{\no}{\noindent}

\def\bea{\begin{eqnarray}}
\def\eea{\end{eqnarray}}
\def\beq{\begin{eqnarray}}
\def\eeq{\end{eqnarray}}

\begin{document}

\title{\textbf{Quantum equivalence between the self-dual and the Maxwell-Chern-Simons
models nonlinearly coupled to $U(1)$ scalar fields }}
\author{D. Dalmazi and Elias L. Mendon\c ca\\
\textit{{UNESP - Campus de Guaratinguet\'a - DFQ} }\\
\textit{{Av. Dr. Ariberto Pereira da Cunha, 333} }\\
\textit{{CEP 12516-410 - Guaratinguet\'a - SP - Brazil.} }\\
\textsf{E-mail: dalmazi@feg.unesp.br }}
\date{\today}
\maketitle

\begin{abstract}

The use of master actions to prove duality at quantum level
becomes cumbersome  if one of the dual fields interacts
nonlinearly with other fields. This is the case of the theory
considered here consisting of  $U(1)$ scalar fields coupled to a
self-dual field through a linear and a quadratic term in the
self-dual field. Integrating perturbatively over the scalar fields
and deriving effective actions for the self-dual and the gauge
field we are able to consistently neglect awkward extra terms
generated via master action and establish quantum duality up to
cubic terms in the coupling constant. The duality holds for the
partition function and some correlation functions. The absence of
ghosts imposes restrictions on the coupling with the scalar
fields.

 \textit{{PACS-No.:}
11.15.-q, 11.10.Kk, 11.10.Gh, 11.10.Ef }
\end{abstract}

\newpage

\section{Introduction}

The use of dual descriptions of the same physical theory is an
important tool in physics as in the AdS/CFT correspondence
\cite{JM}. Deep non-perturbative effects like confinement can be
revealed  \cite{SW} by means of duality. The usual weak coupling
expansion of one theory can describe  the strong coupling regime
of the dual theory and vice-versa as in the case of the massive
Thirring and the Sine-Gordon models in 1+1 dimensions
\cite{Coleman,Abdalla}. In this specific case a theory with at
most quartic interaction is related to a highly nonlinear theory
with all powers of interacting terms. This is in fact similar to
the case discussed in the present work which has its roots in the
duality between the second order Maxwell-Chern-Simons (MCS) gauge
theory and the first order self-dual (SD) model \cite{TPN}.
Although the equivalence between these two free theories, proved
in \cite{DJ} through a master action approach, is interesting in
itself the most powerful applications of duality are found in
interacting theories. It is therefore natural to extend the MCS/SD
duality to include matter interactions \cite{GMS,Anacleto2,jpa},
non-abelian gauge symmetries
\cite{DJ,Karlhede,Bralic,Wotsazek,Botta2}, as well as,
non-commutativity \cite{NC}. In particular, we are interested here
in the coupling of the self-dual field with U(1) charged matter
and its dual gauge theory. The authors of \cite{GMS} have shown
that the gauge theory dual to  U(1) fermions minimally coupled to
the self-dual field must contain a Thirring current-current term
and the minimal coupling has to be  replaced by a Pauli-like
coupling in the dual gauge theory. The proof, based on a master
action, holds for the equations of motion and the partition
function. In \cite{Anacleto2} the case of both charged fermions
and scalar fields minimally coupled to the self-dual field were
considered only at classical level. In the case of scalar fields,
which is considered here, we have an extra complication. Namely,
the dual gauge theory contains besides a Thirring term a highly
nonlinear interaction between the gauge and the matter fields
through the coefficient of the Maxwell term which contains scalar
fields in its denominator. The source of complicated nonlinear
terms is the dependence of the U(1) current on the self-dual field
which is absent for fermions. In \cite{jpa} we have argued that
due to the lack of gauge symmetry in the self-dual model there is
no need for a minimal coupling with the matter fields. Thus, we
can suppress the field dependent part of the U(1) current and work
with a linear coupling in the self-dual field similarly to the
case of fermions where the minimal and linear couplings are the
same. In this case we have been able \cite{jpa} to derive the dual
equivalent gauge theory through a master action and prove the dual
equivalence at quantum level. That corresponds in our notation to
the case $a=0$, see (\ref{lmaster}) and (\ref{mu2}), where the
highly non-linear terms present in the dual gauge theory
(\ref{lmcsm}) disappear. The aim of this work is to return to the
general case $a\ne 0$ and prove the dual equivalence between U(1)
scalar fields nonlinearly coupled to the self-dual field and its
dual gauge theory at quantum level. By calculating  the functional
determinant  from the integral over the scalar fields until
quadratic terms in the coupling we will prove the dual equivalence
at, to that order, of the partition functions and some correlators
thus going beyond the proof of classical equivalence  for
arbitrary values of $a$ given in \cite{Anacleto2,jpa} and quantum
equivalence for $a=0$ presented in \cite{jpa}. Our result is a
nontrivial check of the field dependence of the coefficient of the
Maxwell term appearing in the dual gauge theory.

In subsection 2.1, starting from a master action we recall the
classical equivalence of the self-dual model nonlinearly coupled
to U(1) scalar fields with its dual gauge theory. In 2.2 by
integrating over the matter fields perturbatively, we prove the
dual equivalence of the corresponding partition functions
disregarding cubic and higher terms in the coupling constant. In
2.3 we include sources and extend the proof to correlation
functions. In section 3 we analyze the spectrum of the effective
action for the self-dual field regarding  the presence of ghosts.
At the final section we present the conclusions.

\section{Dual equivalence}

\subsection{Equations of Motion}

Our starting point is the master Lagrangian suggested in
\cite{Anacleto2,jpa}:

\bea {\cal L}_{Master} \, &=& \, \frac{\mu^2}2 f^{\mu}f_{\mu} -
\frac m2\epsilon_{\alpha\beta\gamma}
f^{\alpha}\partial^{\beta}f^{\gamma} - e\, f^{\nu}J_{\nu}^{(0)} +
{\cal
L}_{{\rm Matter}} \nn \\
&+& \frac m2 \epsilon_{\alpha\beta\gamma}
(f^{\alpha}-A^{\alpha})\partial^{\beta}(f^{\gamma}-A^{\gamma}) \label{lmaster} \eea

\no Where

\bea J_{\nu}^{(0)} \, &=& \, i \left(\phi^*\partial_{\nu}\phi -
\phi\partial_{\nu}\phi^*\right)
\label{j}\\
\mu^2 \, &=& \, m^2 \, + \, 2 a e^2 \phi^*\phi \label{mu2} \\
{\cal L}_{{\rm Matter}} \, &=& \, - \phi^*\left(\Box +
m_{\phi}^2\right)\phi\label{lmatter}\eea

\no   We assume
 $g_{\mu\nu}=(+,-,-)$. The quantity $a$ is a
constant and the case of minimal coupling corresponds to $a=-1$.
Since the gauge invariance of the master Lagrangian is guaranteed
for any value of $a$ we do not need to stick to the minimal
coupling. We have shown in \cite{jpa} that from the equations of
motion $\delta {\cal L}_{{\rm Master}} =0$ we can derive two sets
of equations of motion $\delta {\cal L}_{M+SD} =0$ and $\delta
{\cal L}_{M+MCS}=0$ where

\bea {\cal L}_{M+SD} \, &=& \, \frac{\mu^2}2 f^{\mu}f_{\mu} -
\frac m2\epsilon_{\alpha\beta\gamma}
f^{\alpha}\partial^{\beta}f^{\gamma} - e\, f^{\nu}J_{\nu}^{(0)} +
{\cal L}_{{\rm Matter}} \label{lsdm}\eea

\bea \lmmcs = &-&\frac{m^2}{4\mu^2}F_{\alpha\beta}(A)F^{\alpha\beta}(A) + \frac m2
\epsilon_{\alpha\beta\gamma}A^{\alpha}\partial^{\beta}A^{\gamma} \nn \\
&-& \frac{m \, e}{\mu^2}\jo
\epsilon^{\nu\alpha\beta}\partial_{\alpha}A_{\beta} - \frac{e^2}{2
\mu^2}\jo J^{\nu \, (0)} + {\cal L}_{{\rm matter}} \, .
\label{lmcsm} \eea

\no Furthermore, the equations of motion of ${\cal L}_{M+SD}$ and
${\cal L}_{M+MCS}$ are equivalent  to each other through the dual
map $f_{\nu} \leftrightarrow \tilde{A}_{\nu}$ where

\be \tilde{A}_{\mu} \equiv -
\frac{m}{\mu^2}\epsilon_{\mu\nu\alpha}\partial^{\alpha}A^{\nu} +
\frac e{\mu^2} J_{\mu}^{(0)} \label{duality} \ee

\no The classical equivalence holds for arbitrary values of $a$. Concerning the role
of the minimal coupling $(a=-1)$ a comment is in order. Namely, the equations of
motion of $\lmsd$ lead to $\p_{\nu}\left\lbrace\left\lbrack m^2 + \, 2 (a + 1) e^2
\phi^*\phi\right\rbrack f^{\nu}\right\rbrace = 0 $ which works like a gauge condition
assuring that the gauge field $A_{\mu}$ and the self-dual field $f_{\mu}$ have the
same number of degrees of freedom for arbitrary values of $a$. Though it is not
mandatory to fix $a=-1$, in that case we deduce the simple equation $\p_{\nu}f^{\nu}
=0 $ which appears in the free self-dual model.

\subsection{Effective actions}

In order to check
 duality at quantum level
we start with the partition function :

\be {\cal Z} \, = \, \int  {\cal D}\phi{\cal D}\phi^* {\cal
D}f^{\nu} {\cal D}A^{\nu} \, e^{\, \imath\int d^3x\, \lm } \quad .
\label{zmaster1} \ee

\no The gauge field $\, A_{\nu}\, $, after a translation $A_{\nu} \to A_{\nu} +
f_{\nu}$, can be easily integrated leading to:

\be {\cal Z} \, = \, C\, \int  {\cal D}\phi {\cal D}\phi^* {\cal
D}f^{\nu} \, e^{\, \imath\int d^3x\, \lmsd } \quad .
\label{zmaster2} \ee

\no where $C$ is a constant. On the other hand, starting from (\ref{zmaster1}) and
performing the translation $\, f_{\nu}\to f_{\nu} + \left(e\jo - m
\epsilon_{\nu\alpha\beta}\partial ^{\beta}A^{\alpha}\right)/\mu^2 \, $ we arrive at
the dual theory $\lmmcs$ plus an extra term,

 \be {\cal Z} \, = \, \int {\cal D}\phi {\cal D}\phi^*
 {\cal D}f^{\nu} {\cal D}A^{\nu} \, e^{\imath\int
d^3x\left\lbrack \lmmcs + {\cal L}_{{\rm extra }} \right \rbrack } \quad ,
\label{zextra} \ee

\no where

 \be {\cal L}_{{\rm extra }} \, = \, \left(m^2 \, + \, 2 a\, e^2 \,\phi^*\phi
 \right) \frac{f^{\nu}f_{\nu}}2 \quad .
\label{lextra} \ee

\no At classical level, ${\cal L}_{{\rm extra }}$ can be dropped
since its equations of motion imply $f_{\nu}=0$. At quantum level,
the functional integral over $f_{\nu}$ will be matter field
dependent for $a \ne 0$ and there seems to be no simple way to
disregard those potentially divergent contributions. In order to
avoid such problems we have assumed in \cite{jpa} the linear
coupling condition $a=0$ which allowed us to rigorously prove the
dual equivalence between $\lmmcs $ and $\lmsd$ at quantum level
including matter and vector field correlation functions. If $a=0$
we have $\mu^2=m^2$ and the complicated nonlinearities appearing
in (\ref{lmcsm}) disappear. For $a\ne 0$ we need perturbative
methods. Integrating over the scalar fields in (\ref{zmaster1}) we
have\footnote{Throughout this work a small coupling expansion is
understood as an expansion in the dimensionless constant
$e^2/m_{\phi}$} :

\bea {\cal Z} &=& \int  {\cal D}f^{\nu} {\cal D}A^{\nu} \, e^{
\imath\int d^3x {\cal L}(A,f) - {\rm Tr}\ln \left\lbrack -\Box -
m_{\phi}^2 - i\, e\, \left(\p_{\nu}f^{\nu} + 2
f^{\nu}\p_{\nu}\right) + a\,  e^2\,
f^{\alpha}f_{\alpha}\right\rbrack
}\nn \\
&=& \int  {\cal D}f^{\nu} {\cal D}A^{\nu} \, e^{\left\lbrack\,
\imath\int d^3x\, {\cal L}(A,f) + \frac 12\int d^3k f^{\alpha}(-k)
T_{\alpha\beta} f^{\beta}(k) + {\cal O}\left( e^3 \right)
\right\rbrack } \label{zmaster3} \eea

\no where

\be {\cal L}(A,f)  =  \frac{\mu^2}2 f^{\mu}f_{\mu} - \frac
m2\epsilon_{\alpha\beta\gamma} f^{\alpha}\partial^{\beta}f^{\gamma} + \frac m2
\epsilon_{\alpha\beta\gamma}
(f^{\alpha}-A^{\alpha})\partial^{\beta}(f^{\gamma}-A^{\gamma}) \label{laf} \ee

\no The second term in the exponential is written in momentum
space in terms of the Fourier transforms $f_{\nu}(k)$. At
quadratic order in coupling we have only two Feynman integrals. A
careful derivation leads to

\be T_{\alpha\beta}\, = \,  2 a e^2  g_{\alpha\beta} I^{(1)} \, + \, e^2
I_{\alpha\beta}^{(2)} \label{tab} \ee

\no Using dimensional regularization we have obtained for the Feynman integrals:

\be I^{(1)} = \int\frac{d^3 p}{(2\pi)^3}\frac 1{p^2-m_{\phi}^2} = i\frac
{m_{\phi}}{4\pi} \label{i1}\ee

\be I^{(2)}_{\alpha\beta} \!=\! \int \frac{d^3 p}{(2\pi)^3}\frac {(2p + k)_{\alpha}(2p
+ k)_{\beta}} {(p^2-m_{\phi}^2)\left\lbrack (p-k)^2 - m_{\phi}^2\right\rbrack } =
\frac{i m_{\phi}}{8\pi}\left\lbrack 4 g_{\alpha\beta} - \theta_{\alpha\beta}\left( 2 +
\frac{z-1}{\sqrt{z}}\ln \frac{1 + \sqrt{z}}{1 - \sqrt{z}} \right)\right\rbrack \\
\label{i2}\ee

\no With $z=k^2/4 m_{\phi}^2$ and $\theta_{\alpha\beta} =
g_{\alpha\beta} - k_{\alpha}k_{\beta}/k^2$. Our results for
$I^{(1)},I^{(2)}$ are in agreement with \cite{AGS}. The
expressions for the integrals were given in the region $0 \le z
\le 1$. For $z < 0$ we can analytically continue the expressions.
Above the pair creation threshold $( z > 1 )$ the integral
$I^{(2)}_{\alpha\beta}$ develops a real part which will be
neglected here. This is a good approximation for large $m_{\phi}$.
Notice that for $a=-1$ (minimal coupling) $T_{\alpha\beta}$
becomes the vacuum polarization tensor $\Pi_{\alpha\beta}$ of
scalar QED in 2+1 dimensions which is transverse
$k^{\alpha}\Pi_{\alpha\beta}=0=\Pi_{\alpha\beta}k^{\beta}$. Back
in (\ref{zmaster3}) we can write down the effective master action
at quadratic order:

\bea {\cal L}_{{\rm Master}}\, &=& \, \frac{m^2 + c_2}2
f^{\alpha}f_{\alpha} - \frac m2
\epsilon_{\alpha\beta\gamma}f^{\alpha}\p^{\beta} f^{\gamma} -
\frac{c_1}4
F_{\alpha\beta}(f)B(\Box)F^{\alpha\beta}(f) \\
&+& \frac m2
\epsilon_{\alpha\beta\gamma}(A-f)^{\alpha}\p^{\beta}(A-f)^{\gamma}
+ {\cal O}\left( e^3 \right) \label{lmastere2} \eea

\no where we have defined:

\bea  c_1 &=& \frac{e^2}{16\pi m_{\phi}} \label{c1}\\
 c_2 &=& \frac{(a+1)}{2\pi}e^2 m_{\phi} \label{c2}\\
B(\Box) &=& \frac 1z \left( 1 +  \frac{z-1}{2\sqrt{z}}\ln \frac{1 + \sqrt{z}}{1 -
\sqrt{z}} \right) \qquad ; \qquad \label{b} \eea

 Since the quadratic terms in the self-dual field in  (\ref{lmastere2}) are scalar field independent, the
integration over $f_{\nu}$ does not generate unwanted extra terms
as before and we derive, after expanding in the coupling, the
non-local MCS theory:

\be {\cal L}_{{\rm NL-MCS}}^{(e^2)} =  - \frac m2 A^{\mu}\emn
A^{\nu} + \frac{1}4 F_{\mu\nu}(A)\left\lbrack -1 + c_2 + c_1 \Box
B(\Box)\right\rbrack F^{\mu\nu}(A) \label{mcsmnf} \ee

\no On the other hand, if we believe that (\ref{lmcsm}) is the
correct dual gauge theory obtained from the integration over
$f_{\nu}$ in (\ref{zmaster1}), neglecting the extra term
(\ref{lextra}), then it should be possible to derive
(\ref{mcsmnf}) directly from (\ref{lmcsm}) by integrating over the
scalar fields to the quadratic order in the coupling. If we
restrict $\lmmcs$ to the same order $e^2$ and introduce an
auxiliary vector field $B_{\nu}$ to lower the non-linearity of the
Thirring term, the partition function associated with
(\ref{lmcsm}) will be given by

\be {\cal Z}_{{\rm M + MCS}} \, = \, \int  {\cal D}\phi {\cal
D}\phi^* {\cal D}B^{\nu} {\cal D}A^{\nu} \, e^{\, \imath\int
d^3x\, {\cal L}_{{\rm M + MCS}}^{(e^2)}} \quad . \label{zmcsme2}
\ee

 \no where

\bea {\cal L}_{{\rm M + MCS}}^{(e^2)} &=& - \frac m2 A^{\mu}\emn
A^{\nu} + F_{\mu\nu}(A)F^{\mu\nu}(A) \left( - \frac{1}4 + \frac{a
e^2}{2  m^2} \phi^*\phi
\right) \nn \\
&+& \frac{B^{\nu}B_{\nu}}2 - \frac em  \jo \left(B^{\nu} +
\epsilon^{\nu\alpha\beta}\p_{\alpha}A_{\beta} \right) + {\cal
L}_{{\rm matter}} \label{lmcsme2b} \eea

\no We have expanded the coefficient of the Maxwell term up to the
second order in coupling. Integrating over the scalar fields,
using (\ref{i1}) and (\ref{i2}) and Gaussian integrating over
$B_{\nu}$ we obtain

\be {\cal Z}_{{\rm M + MCS}} \, = \, D \int {\cal D}A^{\nu}
e^{\imath\int d^3 x {\cal L}_{{\rm eff}}} \label{zmmcs} \ee

\no With $D$ being a constant. The effective Lagrangian turns out
to match (\ref{mcsmnf}) after expansion ut to the quadratic order
in the coupling:

\bea {\cal L}_{{\rm eff}}^{(e^2)} &=& - \frac m2 A^{\mu}\emn A^{\nu} +
F_{\mu\nu}(A)\left\lbrack \frac{a e^2 m_{\phi}}{8\pi m^2} - \frac{1/4}{1 + \frac{e^2
m_{\phi}}{2\pi m^2} + \frac{c_1 \Box B(\Box)}{m^2}}\right\rbrack F^{\mu\nu}(A)  + {\cal O}(e^3) \nn\\
&=& {\cal L}_{{\rm NL-MCS}}^{(e^2)} + {\cal O}(e^3) \label{match}
\eea

\no Therefore, using (\ref{zmaster1}),(\ref{zmaster2}) and
(\ref{zmcsme2}),(\ref{zmmcs}),(\ref {match}) we have shown that
the partition functions corresponding to the classically
equivalent theories $\lmmcs$ and $\lmsd$ are equivalent to the
order $e^2$ up to an overall constant. In other words, the extra
term (\ref{lextra}) can be completely disregarded to the above
order , although $a\ne 0$.

Now we have na interesting remark about the case of $N_f$ flavors
of scalar fields. This case requires $e\to e/\sqrt{N_f}$ in our
starting Lagrangian (\ref{lmaster}) which would imply $\mu^2 \to
m^2 + (2 a e^2/N_f)\sum_{j=1}^{N_f}\phi_j \phi_j^*)$. It is easy
to convince oneself that the integration over the $N_f$ scalar
fields could be done exactly in the limit $N_f\to \infty$
resulting precisely in our quadratic master action
(\ref{lmastere2}). After integration over the self-dual field we
would obtain

\be {\cal L}_{{\rm NL-MCS}}(N_f\to\infty) =  - \frac m2
A^{\mu}\emn A^{\nu} - \frac{m^2}4 F_{\mu\nu}(A)\frac
1{\left\lbrack m^2 + c_2 + c_1 \Box B(\Box)\right\rbrack
}F^{\mu\nu}(A) \label{mcsnf} \ee

\no On the other hand, we should be able to derive the Lagrangian
above starting from the dual gauge theory ${\cal L}_{{\rm M+MCS}}
+ {\cal L}_{{\rm extra}}$, which now contains $1/\mu^2 =
1/\left\lbrack m^2 + (2 a e^2/N_f)\sum_{j=1}^{N_f}\phi_j
\phi_j^*)\right\rbrack $ in front of the Maxwell term, by taking
$N_f \to \infty$. It turns out that this is not trivial since the
term $\sum_{j=1}^{N_f}\phi_j \phi_j^*/N_f$ which appears in
$1/\mu^2$ is {\it a priori} not small at $N_f\to \infty$. Thus,
the duality allows us to carry out a sum in (\ref{lmcsm}) of
infinite terms of the same order in $1/N_f$ which allows an exact
solution of ${\cal L}_{{\rm M+MCS}} + {\cal L}_{{\rm extra}}$ in
the limit $N_f\to\infty$.

\subsection{Correlation Functions}

Returning to the case $N_f=1$, by introducing sources and
comparing correlation functions  we will show that the dual map
$f_{\nu}\leftrightarrow \tilde{A}_{\nu}$ holds at quantum level.
As in \cite{Banerjee2} we add sources for the dual field
$\tilde{A}_{\nu}$, given in (\ref{duality}). Defining ${\cal D}M
\equiv {\cal D}\phi {\cal D}\phi^* {\cal D}f^{\nu} {\cal
D}A^{\nu}$, we deduce:

\bea {\cal Z}(J) &=& \int {\cal D}M \, e^{\, \imath\int d^3x\left\lbrack \lm +
 J^{\nu}\tilde{A}_{\nu} \right\rbrack} = \int {\cal D}M \, e^{\,
\imath\int d^3x\left\lbrack \lm + f_{\nu}J^{\nu} +
\frac{J_{\nu}J^{\nu}}{2\mu^2}\right\rbrack } \label{zmasterjj1} \\
&=& C\, \int {\cal D}\phi {\cal D}\phi^* {\cal D}f^{\nu} e^{\,
\imath\int d^3x\left\lbrack \lmsd + f_{\nu}J^{\nu} +
\frac{J_{\nu}J^{\nu}}{2\mu^2}\right\rbrack } \label{zmasterjj2}
\eea

\no In  (\ref{zmasterjj1}) we have simply made a translation $f_{\nu} \to f_{\nu} +
J_{\nu}/\mu^2$ while to get (\ref{zmasterjj2}) we did $A_{\nu} \to A_{\nu} + f_{\nu}$
and integrated over the gauge field producing the overall constant $C$. Deriving $\ln
{\cal Z}(J) $ with respect to the sources we can prove the following identity for
connected correlation functions :

\be \left\langle \tilde{A}_{\nu_1}(x_1) \cdots \tilde{A}_{\nu_n}(x_n)
\right\rangle_{{\rm Master}} \, = \,  \left\langle f_{\nu_1}(x_1) \cdots
f_{\nu_n}(x_n) \right\rangle_{{\rm SD + M}} + \, {\rm C.\, T.} \label{wi1}\ee

\no Where ${\rm C.\, T.}$ stands for contact terms. For instance, for the two point
functions we have,

\be \left\langle \tilde{A}_{\nu_1}(x_1)\tilde{A}_{\nu_2}(x_2) \right\rangle_{{\rm
Master}} \, = \,  \left\langle f_{\nu_1}(x_1)f_{\nu_2}(x_2) \right\rangle_{{\rm SD +
M}} + \, g_{\nu_1\nu_2} \delta(x_1-x_2)\left\langle \frac 1{\mu^2}\right\rangle_{{\rm
SD + M}} \label{wi2}\ee

\no From (\ref{wi1}) we see that whatever is the gauge theory obtained from the master
action by integration over $f_{\nu}$, the correlation functions of $\tilde{A}_{\nu}$
in such theory will coincide with the correlation functions of $f_{\nu}$ in $\lmsd$
for arbitrary values of $a$ up to contact terms. Due to the difficulties related with
the integration over $f_{\nu}$ , see (\ref{zextra}), we have to stick once again to
perturbative calculations in order to relate the left handed side of (\ref{wi1}) with
the theory (\ref{lmcsm}). By repeating the steps which have led us from
(\ref{zmaster1}) to (\ref{mcsmnf}) now in the presence of sources we have

\be \int {\cal D}f^{\nu}{\cal D}\phi {\cal D}\phi^* e^{ \imath\int d^3x\left\lbrack
\lm +
 J^{\nu}\tilde{A}_{\nu} \right\rbrack } = e^{\imath\int d^3x {\cal
 L}^{(e^2)}(J)} \label{identity1} \ee

\no Where

\bea {\cal L}^{(e^2)}(J) \, &=& \, {\cal L}_{{\rm NL-MCS}}^{(e^2)} \, + \,
J^{\mu}\left\lbrack \frac{e^2 m_{\phi}}{4\pi m^4}g_{\mu\nu} + \frac{c_1 \Box
B(\Box)}{2 m^4}\theta_{\mu\nu}\right\rbrack J^{\nu} \nn\\
&+& J^{\mu}\left\lbrack \frac 1m - \frac{e^2 m_{\phi} (a+1)}{2\pi m^3} + \frac{c_1
\Box B(\Box)}{m^3} \right\rbrack \epsilon_{\mu\alpha\nu}\p^{\alpha}A^{\nu} + {\cal
O}(e^3) \label{lj}\eea

\no In the expression (\ref{lj}) we have used $\theta_{\alpha\beta}= g_{\alpha\beta} -
\p_{\alpha}\p_{\beta}/\Box $. On the other hand, integrating over the matter fields
disregarding terms of order ${\cal O}(e^3)$, as in the derivation of  (\ref{match})
from (\ref{zmcsme2}), we can deduce

 \be  = \int {\cal D}\phi {\cal D}\phi^* \, e^{
\imath\int d^3x\left\lbrack \lmmcs^{(e^2)} +
 J^{\nu}\tilde{A}_{\nu} + {\cal O}(e^3)
\right\rbrack } = e^{\imath\int d^3x {\cal
 L}^{(e^2)}(J)} \quad . \label{identity2} \ee

\no  From (\ref{identity1}) and (\ref{identity2}) we derive :

\be \left\langle \tilde{A}_{\nu_1}(x_1) \cdots \tilde{A}_{\nu_n}(x_n)
\right\rangle_{{\rm M + MCS}} \, = \, \left\langle \tilde{A}_{\nu_1}(x_1) \cdots
\tilde{A}_{\nu_n}(x_n) \right\rangle_{{\rm Master}}   + {\cal O}(e^3)\label{wi3}\ee

\no From (\ref{wi1}) and (\ref{wi3}) we conclude:

\be \left\langle \tilde{A}_{\nu_1}(x_1) \cdots \tilde{A}_{\nu_n}(x_n)
\right\rangle_{{\rm M + MCS}} \, = \,  \left\langle f_{\nu_1}(x_1) \cdots
f_{\nu_n}(x_n) \right\rangle_{{\rm SD + M}} + \, {\rm C.\, T.} + {\cal O}(e^3)
\label{wi4}\ee

\no Therefore, the mapping $f_{\nu}\leftrightarrow
\tilde{A}_{\nu}$ also holds at quantum level, at least if we
neglect terms of order $e^3$.

For $a=0$ we have shown in \cite{jpa} that matter field correlators in ${\cal L}_{{\rm
M + MCS}}$ and in ${\cal L}_{{\rm M +SD}}$ are equal since no integration over matter
fields is necessary to go from ${\cal L}_{{\rm M + MCS}}$ to ${\cal L}_{{\rm M +SD}}$
via master action. For $a\ne 0$, had we added scalar field sources in
(\ref{zmasterjj1}), that is, instead of $J^{\nu}\tilde{A}_{\nu}$ we had
$J^{\nu}\tilde{A}_{\nu} + \psi \phi + \psi^*\phi^*$, since no scalar field integration
is carried out to obtain (\ref{zmasterjj2}), we would be able to prove that scalar
correlators in ${\cal L}_{{\rm Master}}$ and in ${\cal L}_{{\rm M +SD}}$ would be
equal  which is the analogous of (\ref{wi1}) for pure scalar field correlators.
However, since the matter fields are integrated over perturbatively in
(\ref{identity1}) and (\ref{identity2}), the reader can check that such integral in
the presence of the sources $\psi,\psi^*$ would generate terms of the type $ \psi
(\Box + m_{\phi}^2)^{-1}a e^2 f^2\psi* $ thus leading to divergences in the integral
over the self-dual field which has a delta function propagator, as commented in
\cite{jpa}. Such divergences would invalidate our perturbative integation over the
scalar fields. Therefore, the connection between the scalar field correlators in the
${\cal L}_{{\rm Master}}$ and those correlators in ${\cal L}_{{\rm M + MCS}}$ is more
complicated and we are not able to prove equivalence with the corresponding
correlators in ${\cal L}_{{\rm M + SD}}$ , not even at quadratic order in the
coupling.

\section{Spectrum}

\no After a translation $A_{\nu} \to A_{\nu} + f_{\nu}$ in
(\ref{lmastere2}) we can integrate over the gauge field yielding
an effective non-local self-dual model:

\be {\cal L}_{{\rm NL-SD}} \, = \, \frac{m^2 + c_2}2
f^{\alpha}f_{\alpha} - \frac m2
\epsilon_{\alpha\beta\gamma}f^{\alpha}\p^{\beta} f^{\gamma} -
\frac{c_1}4 F_{\alpha\beta}(f)B(\Box)F^{\alpha\beta}(f)
\label{lff}\ee

\no The effect of the matter fields determinant, up to the
considered order, was to produce another mass term for the
self-dual field plus a non-local Maxwell term.

Now in order to verify whether our quadratic truncation furnishes
sensible theories we check the spectrum of both quadratic theories
(\ref{lff}) and (\ref{mcsmnf}). It is a general result, see
\cite{jhep}, that due to the fact that (\ref{lff}) and
(\ref{mcsmnf}) are connected via a Chern-Simons mixing term, see
(\ref{lmastere2}), the propagators coming from both theories will
have the same pole structure except for a non-physical, gauge
dependent, massless pole $k^2=0$ associated with the Chern-Simons
term which will appear in the propagator of the gauge field as one
can explicitly check from (\ref{mcsmnf}). Consequently, we only
need to check the spectrum of (\ref{lff}). In the large mass limit
$m_{\phi}\to\infty$ ($z \to 0$) using a derivative expansion
$B(\Box) = 2/3 + {\cal O}(-\Box /m_{\phi}^2) $ we recover a local
theory of the Maxwell-Chern-Simons-Proca type:

\bea {\cal L}_{{\rm NL-SD}}^{(e^2)} \, &=& \, \left\lbrack m^2 + \frac{(a+1)}{2\pi}e^2
m_{\phi}\right\rbrack \frac{f^{\alpha}f_{\alpha}}2 - \frac m2
\epsilon_{\alpha\beta\gamma}f^{\alpha}\p^{\beta} f^{\gamma} \nn\\
&-& \frac{e^2}{96\pi m_{\phi}} F_{\alpha\beta}(f)F^{\alpha\beta}(f) + {\cal
O}\left(\frac 1{m_{\phi}^3}\right) \label{mcsproca}\eea

\no It is possible to show \cite{Oswaldo,jhep} that the
Maxwell-Chern-Simons-Proca theory is free of ghosts whenever the
coefficient of the Maxwell term is non-positive and the
coefficient of the Proca term is non-negative. This requires $a
\ge a^* \equiv -1 -(2\pi m^2)/(e^2 m_{\phi})$ which includes the
linear coupling $a=0$ and the minimal coupling $a=-1$. If the
condition $a \ge a^*$ is satisfied we have a perfectly well
defined theory with two massive physical poles. We notice that for
$a\ne -1$ the limit $m_{\phi}\to\infty$ only makes sense if we
assume the scaling $e^2\sim \alpha/m_{\phi}$  where $\alpha$ is
some constant with mass square dimension, after which ${\cal
L}_{{\rm NL-SD}}^{(e^2)}$ becomes a self-dual model with a
modified mass due to the matter fields determinant. At leading
order the Maxwell term is neglected and we end up with just one
massive pole if $a \ne a^*$. In the case $a=a^*$ we have, quite
surprisingly, a gauge theory. The gauge non-invariance of the
non-minimal coupling with the scalar fields cancels the mass term
of the self-dual model. In this special case the duality relates
two gauge theories. On one hand we have a local MCS theory, see
(\ref{mcsproca}) without the Proca term, on the other hand
(\ref{mcsmnf}) becomes for $c_2=-m^2$ and $B(\Box) = 2/3$ a
non-local MCS Lagrangian: $ (m/2) A^{\mu}\emn A^{\nu} - (6\pi m^2
m_{\phi}/e^2) F_{\alpha\beta}(1/\Box)F^{\alpha\beta}$. In
particular, we have the coupling $e^2/m_{\phi}$ on one side and
$m_{\phi}/e^2$ on the dual side which is typical for dual
theories. For the minimal coupling $a=-1$, in the limit
$m_{\phi}\to\infty$, (\ref{lff}) becomes at leading order a pure
self-dual model with the same original mass as before the coupling
to the scalar field. In summary,  both effective theories
(\ref{lff}) and (\ref{mcsmnf}) are perfectly well defined dual
field theories with the same particle content as far as  $a \ge
a^*$.

\section{Conclusion}

The most useful applications of duality concern interacting
theories. It is specially interesting to connect complicated
non-linear theories with simpler dual models. Here we have shown
how a perturbative  integration over part of the degrees of
freedom can help us to find such connections at quantum level.
Explicitly, we have demonstrated, by integrating the scalar fields
to the order $e^2$, that the classical map $f_{\mu}
\leftrightarrow \tilde{A}_{\mu} $ holds also at quantum level at
least perturbatively up to terms of order ${\cal O}(e^3)$. The
quadratic effective free theories obtained lead to sensible
quantum field theories for a large range of the couplings which
includes the  linear $(a=0)$ and the minimal $(a=-1)$ couplings.
In particular, although there is some simplification at classical
level for the minimal coupling, there are apparently no physical
requirements to force us to assume such coupling at quantum level
to the order examined here.

We remark that one of the difficulties in relating nonlinear theories through a master
action is the presence of extra terms like (\ref{lextra}) which have been consistently
neglected here but can possibly  play a role at higher orders in the coupling constant
which demands the inclusion of higher corrections to the the scalar fields determinant
. A complete proof of quantum duality between SD and MCS theories non-linearly coupled
do $U(1)$ scalar fields requires perhaps a non-perturbative analysis of the matter
correlators in both theories. It is tempting to blame the bad infrared behavior of the
self dual field for the infinities related with the extra term (\ref{lextra}).
 The non-abelian and the non-commutative cases of
SD/MCS duality
 suffer from problems alike, i.e., the quadratic terms in the
 self-dual field do not have constant coefficients which makes the
 integral over those fields complicated.

\section{Acknowledgements}

This work was partially supported by \textbf{CNPq}. The work of E.L.M. was supported
by \textbf{FAPESP}. We thank Marcelo Hott for useful discussions.

\end{document}